\documentclass[useAMS]{mn2e}
\usepackage{graphicx}
\usepackage{epsfig}
\usepackage{amssymb}
\usepackage{lscape}
\usepackage{ulem}
\usepackage{txfonts}

\def\kms{km~s$^{-1}$}

\def\c2s{C\,{\sc ii}$^{\star}$}

\def\LIR{L$_{\rm IR}$}

\title[SFRs of AGN] {The star formation rates of active galactic nuclei host galaxies.}

\author[Ellison et al.] {Sara L. Ellison$^1$,
  Hossen Teimoorinia$^1$,
  David J. Rosario$^2$,
J. Trevor Mendel$^2$\\
$^1$ Department of Physics \& Astronomy, University
of Victoria, Finnerty Road, Victoria, British Columbia, V8P 1A1,
Canada.\\
$^2$ Max-Planck-Institut fur Extraterrestrische Physik, Giessenbachstrasse, D-85748 Garching, Germany.
}

\begin{document}

\maketitle

\begin{abstract}
Using artificial neural network (ANN) predictions of total infra-red
luminosities (\LIR), we compare the host galaxy star formation rates (SFRs)
of $\sim$ 21,000 optically selected active galactic nuclei (AGN), 466 low excitation 
radio galaxies (LERGs) and 721 mid-IR selected AGN. SFR offsets ($\Delta$ SFR)
relative to a sample of star-forming `main sequence' galaxies  (matched in M$_{\star}$,
  $z$ and local environment) are computed for the AGN hosts.  Optically
selected AGN exhibit a wide range of $\Delta$SFR, with a distribution skewed to
low SFRs and a median $\Delta$ SFR =  $-0.06$ dex. The LERGs have SFRs
  that are shifted to even lower values 
  with a median $\Delta$ SFR = $-0.5$ dex.  In contrast, mid-IR selected
  AGN have, on average, SFRs enhanced by a factor $\sim$ 1.5.  We interpret
  the different distributions of $\Delta$ SFR amongst the different AGN
  classes in the context of the relative contribution of triggering by galaxy mergers.
  Whereas the LERGs are predominantly fuelled through low accretion rate
  secular processes which are not accompanied by enhancements in SFR,
  mergers, which can simultaneously boost SFRs, most frequently lead to powerful, obscured AGN.
\end{abstract}

\begin{keywords}
  galaxies:  active,  galaxies: Seyfert, galaxies:interactions
\end{keywords}

\section{Introduction}

Nuclear activity is prevalent in star forming galaxies,
indicating that a ready supply of matter for the production of
stars is linked to the fuelling of active galactic nuclei (AGN).
However, the level of star formation in AGN hosts, relative to galaxies
that do not exhibit signs of nuclear activity, is contentious.
Studies of AGN host galaxies have variously shown that they exhibit elevated 
(Silverman et al. 2009; Santini et al. 2012; Juneau et al. 2013;
Rosario et al. 2015; Magliocchetti et al. 2016), normal (Hatziminaoglou 
et al. 2010; Harrison et al. 2012; Rosario et al. 2013;
Stanley et al. 2015; Xu et al. 2015; Lanzuisi et al. 2015) or suppressed
(Salim et al. 2007; Mullaney et al. 2012; Hardcastle et al. 2013; Gurkan et al. 2015; 
Shimizu et al. 2015; Leslie et al. 2016) levels of star formation,
either through `direct' measures of the star formation rate, SFR, or
indirectly through galaxy colour.

At least part of the apparent tension in the literature may be associated
with observational methods and biases in the data.
For example, since star formation rate is known to correlate
closely with stellar mass, and AGN tend be more prevalent in higher
mass galaxies, a simple comparison of AGN and non-AGN will be
biased towards high SFRs in the former.  Comparing SFRs at fixed M$_{\star}$
is therefore essential (e.g. Shimizu et al. 2015; Leslie et al. 2016).
Some disagreement between studies may also arise due to the need to
either stack or average the data in small samples (Mullaney et
al. 2015).  Furthermore, the nature of star formation in AGN
may evolve as a function of redshift (Magliocchetti et al. 2015).
A more subtle point is that AGN may be selected over a broad
range of luminosities and via
a number of diagnostics, spanning the electromagnetic spectrum 
from the X-ray and optical to the mid-infrared (IR) and radio (e.g. see Alexander \&
Hickox 2012 for a review).  AGN selected at different wavelengths
can exhibit very different properties in their accretion
rates, environments and host galaxy properties
(e.g. Kauffmann et al. 2003; Tasse et al. 2008; Hickox et al. 2009; Smolcic  2009;
Best \& Heckman 2012).  The different techniques used
in studies of different AGN samples makes it difficult to fairly
compare the SFRs throughout the AGN family.
A more complete understanding
of the SFRs in AGN therefore requires both large, sensitive samples
for which the distribution of SFRs can be fully parametrized,
and a homogeneous comparison between multi-wavelength selected AGN.

One of the fundamental challenges in the pursuit of a homogeneous,
multi-wavelength assessment of star formation in AGN is the lack
of large samples with uniformly and robustly measured SFRs.
Since the AGN itself `contaminates' the optical recombination
lines that are commonly used to calibrate SFRs, these studies have
had to adopt alternative methods to determine the SFRs.  Two
common approaches have emerged in recent years.
The first method applies the anti-correlation of D$_{4000}$ and specific SFR
measured in star-forming galaxies (e.g. Brinchmann et al. 2004) and
assumes that the same anti-correlation applies in AGN.
The advantage of the D$_{4000}$ SFR calibration is that it
is readily applied to galaxies with optical spectroscopy,
and SFRs can hence be determined for large numbers of local
AGN hosts, such as those in the SDSS.  The main disadvantage of
the D$_{4000}$ technique is that its uncertainties are
relatively large (Rosario et al. 2016).  

More robust SFRs
for local AGN may be determined from far IR luminosities which are
generally considered to be less contaminated by AGN than
other tracers (e.g. Netzer et al. 2007; Buat et al. 2010;
Hatziminaoglou et al. 2010; Mullaney et al. 2011).
However, the use of IR SFR diagnostics to widely characterize
the local galaxy population has so far been relatively limited.
Until recently, the main resource for large area studies of
galactic IR luminosities were the all sky surveys performed
by the Infrared Astronomical Satellite (\textit{IRAS})
(Neugebauer et al. 1984) and \textit{AKARI} (Murakami et al. 2007).
However, these surveys suffer from poor angular resolution
and shallow sensitivity, leading to problems in confusion
and detection of only the highest SFR galaxies.
The \textit{Herschel Space Observatory} (hereafter, simply \textit{Herschel},
Pilbratt et al. 2010),
has recently provided a significant step forward in terms of both
angular resolution and sensitivity.  Several large surveys
have been performed with \textit{Herschel}, such as the \textit{Herschel} Multi-Tiered
Extragalactic Survey (HerMES, Oliver et al. 2012) 
and the \textit{Herschel} Stripe 82 survey (Viero et al. 2014).  However,
even the largest of the \textit{Herschel}  extra-galactic surveys, the
\textit{Herschel} Astrophysical Terahertz Large Area Survey (H-ATLAS, Eales et al.
2010), covered only $\sim$ 550 deg$^2$, which is much smaller
than the areas covered by optical spectroscopic surveys such as the SDSS.
Therefore, a significant
challenge in studying the SFRs of AGN has been assembling samples 
that are both large, and for which robust (e.g. IR) SFRs are available.
Large samples are vital for the measurement of SFR distributions, rather than
binned averages which can be dominated by outliers (Mullaney et al. 2015).
In Ellison et al. (2016) we presented a catalog of $\sim$ 330,000
infra-red luminosities (\LIR) for galaxies in the SDSS, derived
by artificial neural network (ANN) techniques.    In the
current work, we use the \LIR\ catalog of Ellison et al. (2016)
to explore the SFRs of AGN in the local Universe.


\section{Methodology}

Full details of the \LIR\ determinations used in this work are
provided in Ellison et al. (2016); only a brief summary is presented
here.  The \textit{Herschel} Stripe 82 Survey (Viero et al. 2014) covers
a total of 79 deg$^2$ in an equatorial strip of the SDSS footprint.
Rosario et al. (2016) cross-matched the 250 $\mu$m detected galaxies
from \textit{Herschel} with the SDSS DR7 Main Galaxy Sample whose extinction corrected
Petrosian r-band magnitudes are brighter than 17.77 and whose
redshifts are in the range 0.04 $< z <$ 0.15, using a positional
matching tolerance of 5 arcsecs.  The \textit{Herschel}-SDSS matched
catalog contains 3319 galaxies, which were further cross-matched
with the \textit{Wide Field IR Sky Explorer} (\textit{WISE}, Wright et al. 2010)
catalog to yield photometry spanning the mid- to
far-IR.  Infra-red luminosities were derived by Rosario et al. (2016)
by fitting the mid- and far-IR photometry with the spectral energy
distribution templates of Dale \& Helou (2002).  AGN are
expected to contribute negligibly to the \LIR\ of this sample
(see Rosario et al. 2016 for more details).

The \LIR\ is estimated from an artificial neural network that uses
1136 \textit{Herschel} detected galaxies with 23 physical parameters, such
as stellar masses (Mendel et al. 2014), emission line strengths and
photometry (Simard et al. 2011), measured from SDSS data.
Based on the popular conversion presented by Kennicutt (1998),
the \LIR\ may be converted to star formation rates (for a Chabrier
initial mass function) using

\begin{equation}\label{eqn-sfr}
\log  L_{IR} (erg/s) = \log SFR (M_{\odot}/yr) + 43.591.
\end{equation}

The SFRs that result from the ANN predictions were shown by
Ellison et al. (2016) to be
in excellent agreement with those from the SDSS for star-forming
galaxies.  In this paper, we additionally utilize the \LIR\ values of the
Ellison et al. (2016) catalog to determine SFRs for AGN host galaxies;
indeed, \LIR\ SFRs are used  throughout this paper, for both star-forming
and AGN dominated galaxies, unless otherwise stated.  Throughout
this paper we require $\sigma_{\rm ANN}$ (the uncertainty
on the ANN predicted \LIR) to be less than 0.1 dex (247,137 galaxeis) and convert \LIR\
to SFR using equation \ref{eqn-sfr}.  

In order to compare the SFRs of AGN to those of star-forming (non-AGN)
galaxies, we define the SFR offset ($\Delta$ SFR) of a given galaxy
(SFR$_{\rm gal}$) relative to a set of comparison star forming (as 
defined by Kauffmann et al. 2003 and with S/N$>3$) galaxies
that are matched in stellar mass, redshift and local galaxy density
(environment), also selected from the SDSS.  Local density is defined as
$\Sigma_5 = \frac{5}{\pi d_5^2}$,
where $d_5$ is the projected distance in Mpc to the $5^{th}$ nearest
neighbour within $\pm$1000 \kms.  Normalized densities, $\delta_5$,
are computed relative to the median $\Sigma_5$ within a redshift slice
$\pm$ 0.01.  
The baseline tolerance used for matching is 0.1 dex
in M$_{\star}$, 0.005 in $z$ and 0.1 dex in $\delta_5$.
We require at least 5 comparison galaxies in the matched sample;
if this is not achieved then the mass, redshift and local density tolerances
are grown in further increments of 0.1 dex, 0.005 and 0.1 dex respectively,
until the minimum size criterion of 5 is met. 
The SFR of the comparison star forming sample (SFR$_{\rm comp}$) is taken as the median
of the matched control sample. The SFR offset is then defined as:

\begin{equation}
  \Delta SFR = \log SFR_{\rm gal} - \log SFR_{\rm comp}.
\end{equation}

\section{Results}

\subsection{Optically selected AGN}

Baldwin, Phillips \& Terlevich (1981) pioneered the use of emission
line ratios to distinguish galaxies dominated by various photoionizing
processes.  Diagnostics separating star-forming from AGN dominated galaxies
are now commonly referred to as `BPT' diagrams, in reference to these
authors.  In this paper, we make use of three of the most commonly
used diagnostics which incorporate the ratios of [NII]/H$\alpha$
and [OIII]/H$\beta$, by Kewley et al. (2001), Kauffmann et al. (2003) and
Stasinska et al. (2006), hereafter K01, K03 and S06 respectively.
S06 identifies galaxies with even a small AGN contribution.  K03
is slightly more relaxed in its selection of AGN but still includes
galaxies that are composites of star formation and AGN.
AGN identified by K01 are those dominated by the AGN.

\begin{figure}
\centering
\includegraphics[width=9.cm,angle=0]{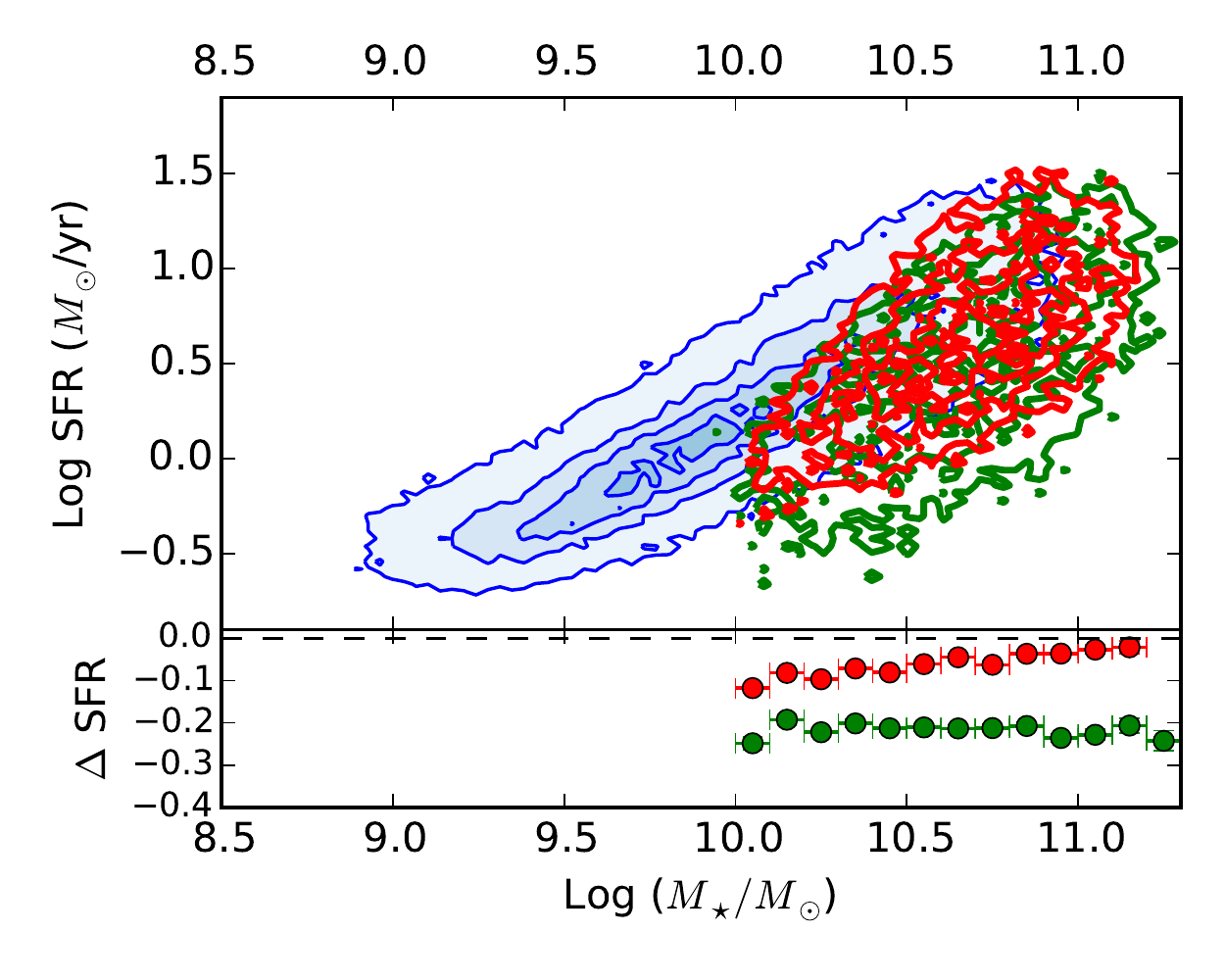}
\caption{Upper panel: A comparison of star formation rates (from ANN \LIR\
   estimates) in star forming galaxies
  (filled blue contours) and AGN (open contours) classified according to Kauffmann
  et al. (2003).  Lower panel: The
  offset between SFRs in AGN (as measured from \LIR) relative to the main
  sequence.  In both panels, red and green colours refer to SFRs derived from
  ANN \LIR\ predictions and from D$_{4000}$ measurements, respectively.  }
\label{MS_K03}
\end{figure}

We begin our analysis with a single definition of `AGN', adopting
K03 as our fiducial classification (although our results are qualitatively
similar with any of the 3 diagnostics), with a S/N$>$5 requirement to minimize the
contribution by Low Ionization Nuclear Emission line Regions (LINERs;
see Leslie et al. 2016 for an assessment of SFR offsets in the LINER population).
These selection criteria yield
20,926 AGN with robust ($\sigma_{\rm ANN} < 0.1$ dex)
\LIR\ predictions.    Fig. \ref{MS_K03} shows the
location of the main sequence of star forming galaxies as filled
blue contours.  The SFRs of AGN derived from the ANN \LIR\ are shown
in red contours.  For comparison, we also show the aperture
corrected AGN SFRs derived from the SDSS spectra via the D$_{4000}$
method.   

As found by previous authors using the D$_{4000}$ method to measure
SFRs (e.g. Salim et al. 2007; Shimizu et al.
2015; Leslie et al. 2016), the optically selected AGN clearly
have \LIR\ SFRs that lie below the main sequence, i.e. AGN have low SFRs
for their M$_{\star}$, typically by 0.05 -- 0.1 dex, with a median $\Delta$ SFR=-0.06.
The main sequence offsets
measured using D$_{4000}$ extended to somewhat lower SFRs than the \LIR\ derived values.
There are at least two possible reasons for this.  First, the
SFRs derived from the ANN are limited to about 0.3 M$_{\odot}$ yr$^{-1}$,
due to the detection limit in the training set.  It is therefore
likely that there is a further tail of low SFR AGN that are not included
in our sample due to an effective `detection limit' in the ANN
\LIR\ predictions (e.g. Leslie et al. 2016).
A second possible reason for the more extended green contours is that, as
shown by Rosario et al. (2016) and Ellison et al. (2016), the
D$_{4000}$ based SFRs are much more uncertain than those derived
from \LIR, so the lower SFRs in the former may be a result of
a broader error disribution.  

\begin{figure}
\centering
\includegraphics[width=9.cm,angle=0]{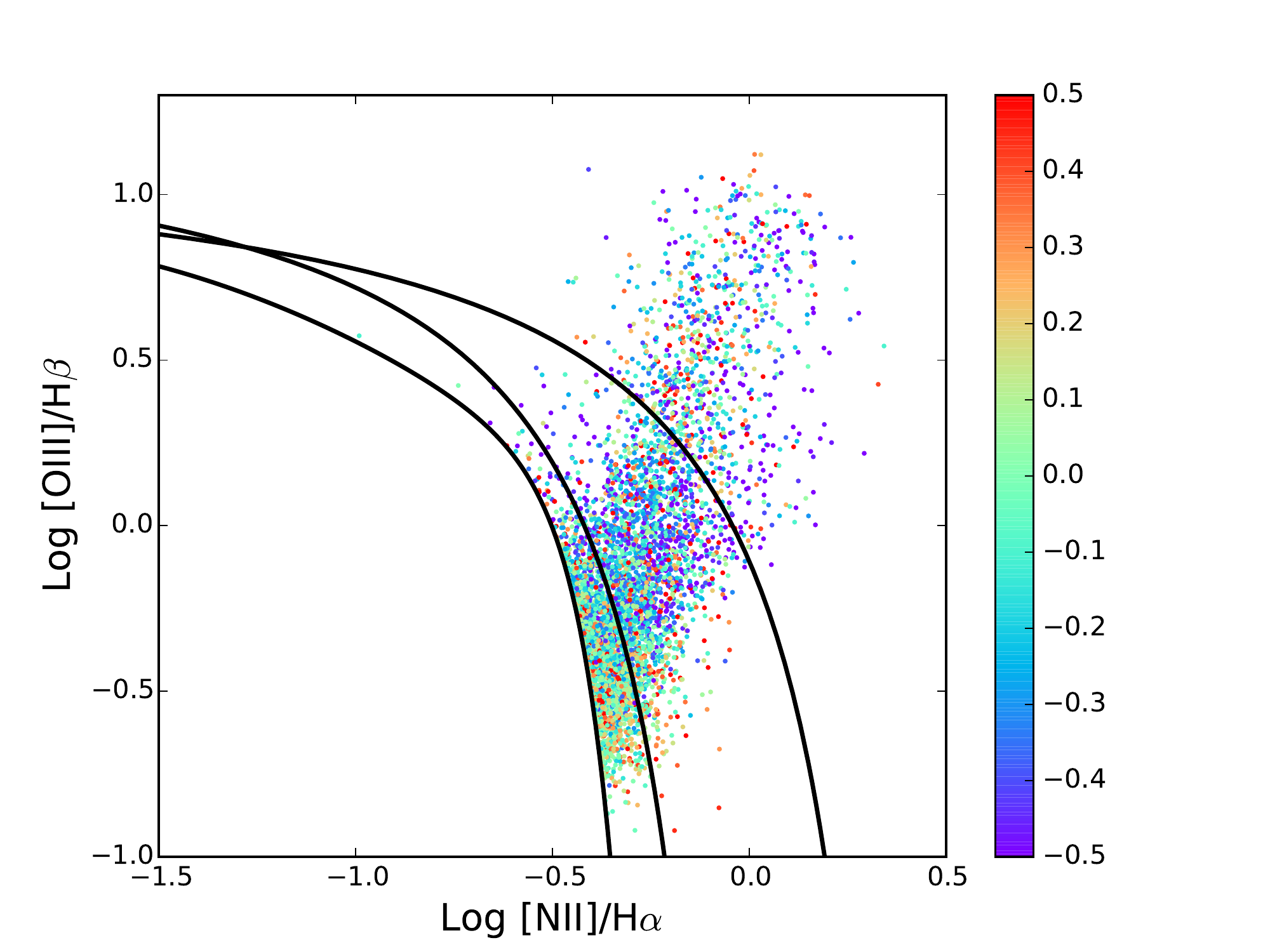}
\caption{BPT diagram colour coded by $\Delta$ SFR for AGN.The 3 black lines
  show the AGN demarcations of (from top to bottom), K01, K03 and S06.
  A random sample of 5000 out of $\sim$33,300 AGN selected by S06 with
  \LIR\ based SFRs is shown.  }
\label{BPT_dsfr}
\end{figure}

In order to investigate the dependence of $\Delta$ SFR on the relative
contribution of the AGN, in Fig. \ref{BPT_dsfr} we plot the BPT
diagram for AGN colour-coded by $\Delta$ SFR.   In order to explore the
dependence of $\Delta$ SFR over the maximum range on the BPT diagram,
we `relax' the definition of AGN by adopting the criteria of S06, but also
showing the demarcation lines fo K01 and K03.   There are $\sim$ 33,300 galaxies that are classified
as AGN by the S06 criterion for which we have ANN \LIR\ predictions and
for which $\Delta$ SFR can be determined.  To avoid crowding on Fig.
\ref{BPT_dsfr} we plot a random sample of 5000 AGN.
For reference, all three commonly used diagnostics (K01, K03, S06)
are shown in Fig. \ref{BPT_dsfr} as black lines.
Negative values of $\Delta$ SFR (low SFRs) dominate throughout the AGN `wing'.
The exception is very close to the
S06 demarcation, where there appear to be fewer galaxies with suppressed SFRs.
Despite the general preference for AGN to exhibit low $\Delta$ SFRs,
there is a wide variety of
SFR enhancements at most locations.  Fig. \ref{BPT_dsfr}
demonstrates that, although there is a trend towards more suppressed SFRs as
we move along the AGN branch (see also Leslie et al. 2016), even galaxies
that are dominated by AGN can sometimes exhibit strong SFR enhancements
(see also the middle panel of Fig. \ref{all_dsfr}).  An important caveat 
to these conclusions is that samples of optically selected  AGN may
be inherently biased against galaxies with very high SFRs.  Trump
et al. (2015) have investigated the effect of star formation `dilution',
whereby high SFRs can move even moderately powerful AGN onto the
star forming branch of the BPT diagram.  The dilution effect is most
problematic for low mass galaxies.  However, as shown in the lower
panel of Fig. \ref{MS_K03} the median value of $\Delta$ SFR in optically
selected AGN does not depend strongly on M$_{\star}$, indicating that
any bias against the identification of highly star-forming AGN is not
driving the tendency towards suppressed SFRs.   

For the remainder of this paper, we use the K03 classification
as our fiducial definition of `optical' AGN.

\subsection{AGN selected in the IR and radio}

The physical properties, such as stellar mass distribution, morphological
structure and average colours, of AGN host galaxies depend strongly on how the
AGN are selected (e.g. Hickox et al. 2009).  In this sub-section, we determine
SFR offsets from the main sequence for two other samples of AGN, one selected
based on radio emission, and the other based on mid-IR colour selection.
The novelty of our work is that we can make a direct comparison of
SFR offsets for these three populations based on a homogeneous analysis.

The sample of low luminosity radio-selected AGN, specifically
the low excitation radio galaxies (LERGs), is taken from the SDSS catalog
compiled by Best \& Heckman (2012).   We determine the SFRs of the Best \&
Heckman LERG catalog from our ANN LIR predictions; there are 466
LERGs for which $\sigma_{\rm ANN} < 0.1$.  

The sample of mid-IR selected AGN
is identified by cross-matching the \textit{WISE} All Sky Survey (Wright et al. 2010)
with the SDSS and requiring a maximum angular separation of
6 arcsecs.  We use the W1 and W2 band profile magnitudes, with
a minimum S/N requirement of 5 and identify AGN as galaxies exceeding
W1$-$W2$>$0.8 (Stern et al. 2012).  721 of these \textit{WISE} selected AGN have ANN
\LIR\ with $\sigma_{\rm ANN} < 0.1$.  

\begin{figure}
\centering
\includegraphics[width=9.cm,angle=0]{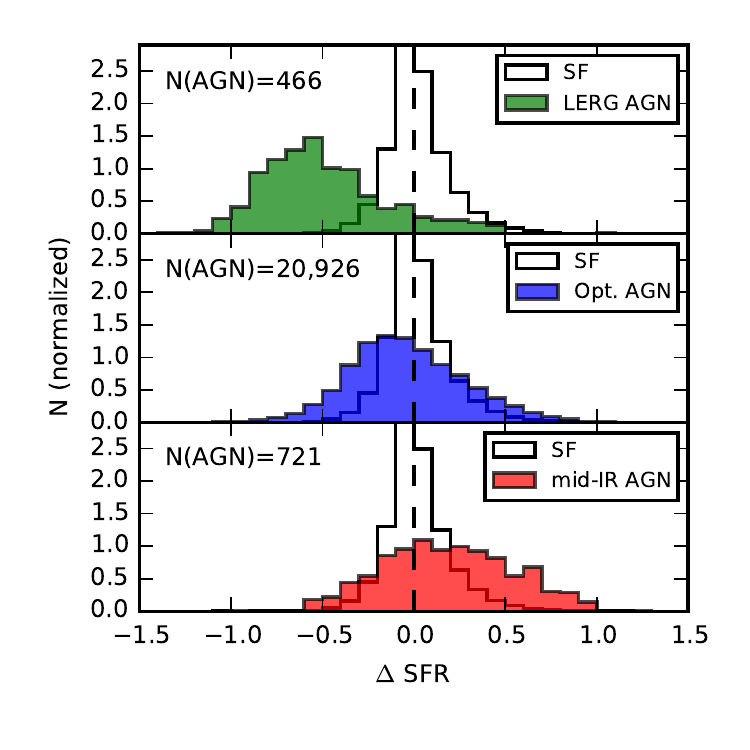}
\caption{Distributions of $\Delta$ SFR for various classes of AGN.
  In all panels, the distribution of $\Delta$ SFR in star forming
  galaxies is shown in black; by definition, this distribution is
centred around $\Delta$ SFR = 0.  }
\label{all_dsfr}
\end{figure}

In Fig. \ref{all_dsfr} we show the SFR offsets of the three
AGN samples (coloured histograms), as well as the distribution
of $\Delta$SFR amongst star-forming galaxies (black line).
 By definition, this latter distribution
should be centred at $\Delta$ SFR = 0, since the star forming
galaxies define the main sequence from which offsets are computed;
the width of the histogram simply shows the scatter in the main sequence.
In the upper panel of Fig. \ref{all_dsfr} we confirm the result
of Gurkan et al. (2015), who studied LERGs in the H-ATLAS sample,
that LERGs are offset to lower SFRs than the star forming main sequence.  
The optically selected AGN (K03) are shown in the blue
histogram of the middle panel of Fig. \ref{all_dsfr}.
By comparing the top and middle
panels  of Fig. \ref{all_dsfr} it can be seen that,
whilst both the LERGs and the optically selected AGN have a
modal $\Delta$ SFR that is negative, the
distributions are quite different.  On average, the LERGs are shifted
to much lower values of $\Delta$ SFR than the optically selected
AGN; the median offset is $\Delta$SFR=$-0.5$.  
Whilst 65 per cent of optically selected AGN lie within
a factor of two of the main sequence, and can even exhibit
enhanced SFRs, the LERGs rarely attain the main sequence SFR
for their stellar mass. 

In the lower panel of Fig. \ref{all_dsfr} we show the distribution
of SFR offsets for AGN identified in the mid-IR.  For the first time, we
can now compare the SFR offsets in a homogeneous way for the three
AGN populations. In contrast to the optically selected AGN
and the LERGs, the median $\Delta$ SFR of mid-IR selected AGN
is +0.17 dex, indicating that the typical powerful, dust-obscured AGN has a factor of $\sim$
1.5 excess SFR compared with the main sequence.  However, the tail
of the $\Delta$ SFR distribution extends up to +1 dex, indicating
that some mid-IR selected AGN have SFRs a factor of 10 above their
comparison sample.  Such elevated SFRs are absent in the LERG sample,
and rare amongst the optically selected AGN. In Ellison et al.
(2016) we showed that the ANN predictions of \LIR\ were accurate
even when the SFR was elevated to these levels above the main
sequence.

\section{Discussion and conclusions}

What drives the difference between the $\Delta$ SFR distributions
of the 3 types of AGN shown in Fig. \ref{all_dsfr}?  In particular,
why are the SFRs in mid-IR selected AGN enhanced relative to the main
sequence, whereas LERGs have SFRs that are almost universally
below expectations for their M$_{\star}$ (see also
  Gurkan et al. 2015)? Hardcastle et al. (2013)
posed a similar question after finding that high excitation radio
galaxies (HERGs) had enhanced SFRs, in contrast to the low SFRs
in LERGs.  Hardcastle et al. (2013) proposed this difference was due
to the environments in which LERGs an HERGs are typically located
(Best \& Heckman 2012).  However, since our star-forming comparison
sample is matched in $\delta_5$, this seems unlikely to be the
explanation for the different $\Delta$ SFR distributions seen in
Fig. \ref{all_dsfr}.  

Our finding that IR, optical and radio-selected AGN form a
sequence in $\Delta$ SFR is consistent with an evolutionary picture
in which an abundant gas supply triggers 
enhanced star formation in an obscured AGN phase.  As the fuel supply
declines, so does the star formation rate and the obscuration.
A similar picture has been proposed by Hickox et al. (2009),
based on a comparison between the host galaxy and clustering 
properties of a sample of IR, X-ray and radio selected AGN.
Hickox et al. (2009) found that AGN selected in these ways
formed a sequence in decreasing star formation and accretion rate,
and increasing halo mass.
The evolutionary sequence between star formation and obscuration
is further supported by the observation that, for samples of X-ray selected
AGN, star formation rates are significantly enhanced for
the obscured/absorbed AGN, but are normal or suppressed
for unobscured AGN (Stevens et al. 2005; Juneau et al. 2013).
Moreover, the X-ray absorbed AGN fraction increases as
galaxies lie progressively above the star forming main
sequence (Juneau et al. 2013).

Galaxy mergers are a natural mechanism to explain such an 
evolutionary sequence, as first proposed by Sanders et al. (1988),
since mergers are known to be capable of triggering both star
formation (e.g. Scudder et al. 2012) and AGN activity (e.g. Ellison
et al. 2011).  In this scenario, the merger remnant of a gas-rich, 
major merger first undergoes a significant
starburst, possibly manifesting as a luminous (or ultra-luminous) IR
galaxy and high accretion rate AGN (possibly a quasar).  As the starburst
fades, the AGN transitions from obscured to unobscured and the
galaxy eventually evolves to an early type (e.g. see Fig. 6 in
Alexander \& Hickox 2012 for a schematic view of this process).

The dependence of star formation rate enhancement on AGN obscuration 
during the post-merger phase is supported by the observation that
certain classes of AGN are more prevalent amongst ongoing mergers.
For example, in close pairs and recent post-mergers, optically
  selected AGN are over-abundant (relative to the AGN frequency
  in isolated galaxies) by a factor of 2--3 (Ellison et al. 2013),
compared to a factor of 5--20 in mid-IR selected AGN
(Satyapal et al. 2014).  Mid-IR selected AGN have
also been shown to exhibit a high dual black hole fraction
(Secrest et al., in preparation), in contrast to optical searches
which have a low success rate in detecting binaries (Fu et al. 2011;
Muller-Sanchez et. al 2015).  The connection between obscuration and
merging is also found in the X-ray, where the most absorbed AGN are
found most frequently in mergers (Kocevski et al. 2015; Lanzuisi et al.
2015).

In contrast to the consistent link between obscured AGN and mergers,
the results for the unobscured population has been more controversial.
Whereas an excess of AGN is measured in close pairs of galaxies (Ellison
et al. 2011; Silverman et al. 2011), demonstrating that mergers
\textit{may} trigger AGN, the morphological classification of AGN
indicates that mergers do not appear to \textit{dominate}
the triggering of unobscured AGN (e.g. Gabor et al. 2009;
Kocevski et al. 2012).  These results are consistent with the
distribution of SFRs in optically selected AGN (Fig \ref{all_dsfr},
middle panel), in which the
majority of galaxies have normal-to-low SFRs, and are hence unlikely
to be linked to mergers.  The small number of optical AGN with enhanced
SFRs may be those that are merger triggered, a postulate that is
confirmed through visual inspection of the SDSS images.

In contrast to optical and mid-IR selected
AGN, there is no excess of LERGs in merger samples, indicating that
they are fuelled by secular processes,
where low levels of accretion (not connected with a starburst)
can be maintained from a combination of
stellar and external sources  (Ellison, Hickox \& Patton 2015).
Rather than an `evolutionary' (i.e. time sequence) scenario,
our observations may therefore also be explained by different relative 
contributions of mergers to the 3 AGN classes.  I.e.
LERGs have the lowest SFRs because
there is no boost from mergers, whereas a higher fraction of IR
selected AGN are linked to mergers and have experienced triggered
star formation during the interaction.  Put another way, mergers
may lead to high SFRs and luminous (obscured) AGN, but a distinct, secular
pathway is responsible for lower luminosity AGN with normal or low
SFRs (e.g. Shao et al. 2010).

The idea that the distribution of $\Delta$ SFR in the different
AGN classes is by driven by the relative contributions of mergers
is supported by differences in their merger fractions.
We compute the merger fraction for each AGN class as the fraction
of galaxies classified as a merger in Galaxy Zoo, requiring a
merger vote fraction $>$ 0.4 (e.g. Darg et al. 2010).
Merger fractions for the different types of AGN are: radio - 1 per cent,
optical - 3 per cent, mid-IR - 7 per cent.  The higher merger fraction
amongst the mid-IR sample is despite a redshift distribution skewed
towards slightly higher values than the optical AGN and
LERG samples making it potentially harder to identify mergers therein.
Although these merger fractions should not
be considered as absolute, due to the limitations of visual classification
in shallow ground-based images, without removal of projected companions,
their relative values are indicative that mergers are most prevalent amongst IR
selected AGN, and least common amongst LERGs.  A complete assessment of
mergers in this sample is beyond the scope of this paper, but we note that the excess
of mergers amongst mid-IR selected persists for stricter cuts on merger
vote fraction.

In conclusion, our main result is that the SFRs of AGN are critically
dependent on how an AGN is selected. IR selected galaxies in the SDSS
show a median SFR enhancement of a factor of $\sim$ 1.5, compared to
an under-abundance of star formation in optically selected AGN (by
$\sim$ 25 per cent) and low luminosity radio-selected AGN (by a factor of
3).  We propose that these differences may be explained by a relatively
high fraction of mergers amongst IR selected AGN, a lower merger
incidence amongst optically selected AGN, and the domination of
secular fuelling processes for LERGs, which are hosted primarily by
massive galaxies with low SFRs.


\end{document}